# Generation of vector beams in planar photonic crystal cavities with multiple missing-hole defects

Chenyang Zhao[†], Xuetao Gan[†], Sheng Liu, Yan Pang, Jianlin Zhao*

*Key Laboratory of Space Applied Physics and Chemistry, Ministry of Education, and Shaanxi Key Laboratory of Optical Information Technology, School of Science, Northwestern Polytechnical University, Xi'an 710072, China*

*† These authors contributed equally to this work.*

*Corresponding author: jlzhao@nwpu.edu.cn*

**Abstract:** We propose a novel method to generate vector beams in planar photonic crystal cavities with multiple missing-hole defects. Simulating the resonant modes in the cavities, we observe that the optical fields in each defect have different phase and polarization state distributions, which promise the compositions of vector beams by the scattered light from the defects. The far-field radiation patterns of the cavity modes calculated via the Sommerfeld diffraction theory present vector beams possessing hollow intensity profiles and polarization singularities. In addition, the extraction efficiencies of the vector beams from the cavities could be improved by modifying the air-holes surrounding the defects. This planar photonic crystal cavity-based vector beam generator may provide useful insights for the on-chip controlling of vector beams in their propagations and interactions with matter.

## 1. Introduction

Vector beams, with spatially inhomogeneous polarization states [1], have attracted growing attentions due to their distinctive properties, including tight-focusing [2-6], orbital angular momentum (OAM) [7-9] and controllable nonlinear dynamics [10]. Such researches provide great application potentials of vector beams in surface plasmon excitation [11], superresolution imaging [12], optical manipulation [13], laser micromachining [14], and so on. Approaches based on diffraction optical elements [15-17] and interferometric techniques [18,19] have been demonstrated to convert fundamental laser modes to vector beams in free-space. On the other hand, generation of vector beams in photonic chips also attract significant interest, as it expands the applications of vector beams into photonic integrated circuits with improved stability and compact footprint. Already, with the fabrications of in-ring angular gratings [8] or surrounded nano-rods onto micro-ring resonators [20], vector beam emitters have been proposed and fabricated on silicon-on-insulator chips. Moreover, the ring-based emitters enable the manipulations of polarization states and OAM of the vector beams by changing the grating structures. More recently, Schulz *et al.* proposed an on-chip quantum computation with the OAM *qudit* states and sorters generated by ring emitters [9].

Planar photonic crystal (PPC), as one of the important chip-integrated photonic architectures, enables the control of light propagation and interaction with matter in wavelength-scale [21-25]. With their controllable photonic band structures, PPCs allow more reliable manipulations on light phases and polarization states via its Bloch modes, band-edge modes, and defect modes [26-28]. For instance, Noda *et al.* reported on-chip lasers of vector beams relying on the slow-light effect at the band-edge of PPCs [29-31]. Generation of vector beams in PPCs is expected to develop vector beam-based on-chip photonic circuits, classical and non-classical light sources, and coupled quantum dot cavity systems. In this paper, we demonstrate the generation of vector beams in PPC cavities by designing multiple missing-hole defects. Different from the vector beams generated from the superposition of Bloch modes of PPCs [29, 30], the cavity-based vector beams are bound states of the PPC defects and have much smaller mode volumes in the wavelength-scale. These attributes indicate more reliable applications in micro-laser and on-chip integrations of linear and nonlinear devices of vector beams.

**2. PPC cavity designs and theoretical analysis**

We study the generation of vector beams in a triangular-lattice PPC structure with lattice constant of $a$, air-hole radius of $r$, slab thickness of $d$, and slab refractive index of $n_{\text{slab}}$. The defect of the PPC cavity is introduced by symmetrically missing $N$ separated holes with respect to the central air-hole, as shown in Fig. 1(a) for the case of $N=6$. In PPC cavities, as a result of the in-plane distributed Bragg reflection and the vertical total internal reflection, resonant modes are confined in the missing-hole defects. Moreover, relying on different geometry orientations of the defects, field distributions in each defect are expected to have different polarization directions with respect to the central air-hole. On the other hand, due to the existence of light components with *k*-vectors in the light cone of the PPC slab, the resonant modes would scatter out via the defects, which is responsible for the main energy-loss. The far-field light $E_j(x, y, z)$ scattered from the *j*th defect can be expressed as

$$\mathbf{E}_j(x,y,z) = \frac{\mathbf{A}_j(x_0,y_0,0)\exp(ikR)}{R} \quad (1)$$

where $k=2\pi/\lambda$ is the wave number, $R=[(x-x_0)^2+(y-y_0)^2+z^2]^{1/2}$ and $j=1, 2, …, N$. $A_j(x_0,y_0,0) \propto E_j(x_0, y_0, 0)\mu_j$, where $E_j(x_0, y_0, 0)$ is the electric field of the near-field resonant mode in the $j$th defect and $\mu_j$ is the unit vector describing the polarization direction. Based on the superposition principle of optical fields, the coherent combination of $E_j(x, y, z)$ scattered from the $N$ defects will present a complex field pattern, which depends on near fields in each defect and their polarization states, as illustrated in Fig. 1(b).

To verify the formation of vector beams from the overall scattered light of resonant modes in the PPC cavity, we assign six dipole oscillations with different polarization states in the center of the cavity defects, as indicated by the red dots in Fig. 1(a), and calculate their interference in the far-field radiation. Figures 1(c) and 1(d) show the superposed optical fields of the discrete dipoles with symmetrically azimuthal (upper) and radial (lower) polarization states relative to the central point, where the left, middle, and right images denote the near-field intensity distribution, far-field radiation patterns calculated using Sommerfeld diffraction theory and COMSOL Multiphysics software, respectively. The results reveal vector beams with azimuthal or radial polarizations are generated, depending on the polarization distributions of the discrete dipoles. In addition, the simulation results using COMSOL Multiphysics software prove the validation of Sommerfeld diffraction theory in the calculations of far-field vector beams.

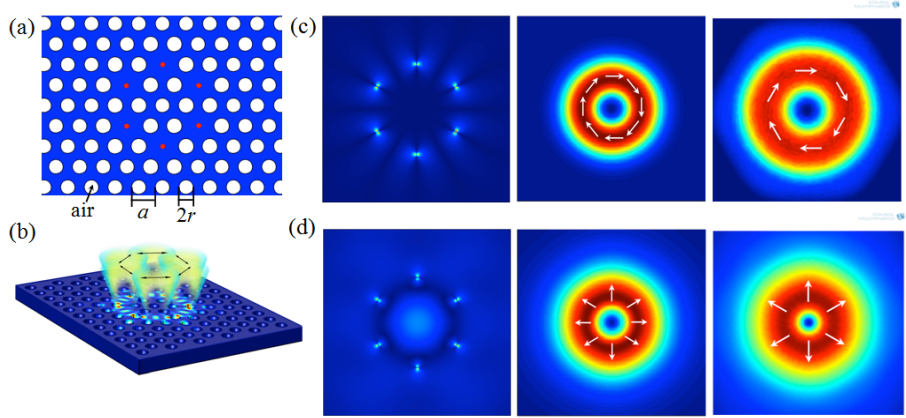

Fig.1. Illustrations of vector beam generation using PPC cavities with multiple point-defects. (a) Air-holes structure of the PPC cavity in the case of $N=6$; (b) schematic illustration of the vector beam formation; (c)-(d) near-field (left) and far-field (middle and right) intensity distributions of six dipoles with azimuthal (upper) and radial (lower) polarization states, where the middle and right images show the results calculated by Sommerfeld diffraction theory and COMSOL Multiphysics software, respectively. The locations of the six dipoles are indicated by the red dots in (a).

## 3. Numerical simulations and discussions

We then demonstrate the feasibility of above analysis by simulating resonant modes of the multi-defect PPC cavities with a three-dimensional finite element mode technique (COMSOL Multiphysics). To provide the reliability of simulation results in realistic applications, the PPC cavities are designed for a gallium arsenide (GaAs) slab with embedded indium gallium arsenide (InGaAs) quantum dots (QDs), which can provide an efficient internal light source [22] and a platform for studying the cavity quantum electrodynamics (QED) [25]. The thickness and refractive index of the GaAs slab are chosen as $d=220$ nm and $n_{slab}=3.46$, respectively. The PPC structure has a lattice constant of $a=280$ nm and an air-hole radius of $r = 0.3a$.

*3.1 Near-field resonant modes in a PPC cavity with N=6*

Simulating a PPC cavity with six point-defects, we obtain two resonant modes at wavelengths of 972 nm (Mode1) and 957 nm (Mode2), locating in the emission range of InGaAs QDs [32]. However, due to the scattering loss from the defects, the quality ($Q$) factors of the two modes are low as 430 and 420, respectively. The large scattering loss, in turn, enables an effective collection of the resonant modes in the far-field. The near-field intensity distributions of Mode1 and Mode2 are depicted in Figs. 2(a1) and 2(a2), respectively, showing strong light fields confined in the defects. The point-defect structure of the cavity enables the mode volumes ($V_{mode}$) to be low as 2.17 and 2.07 ($\lambda/n_{slab})^3$ for Mode1 and Mode2, which could promise strong light-matter interactions in the cavity. In the figures, the superimposed arrows represent the polarization states of the resonant fields, showing inhomogeneously distributed polarizations with respect to the central air-hole, as indicated in the insets.

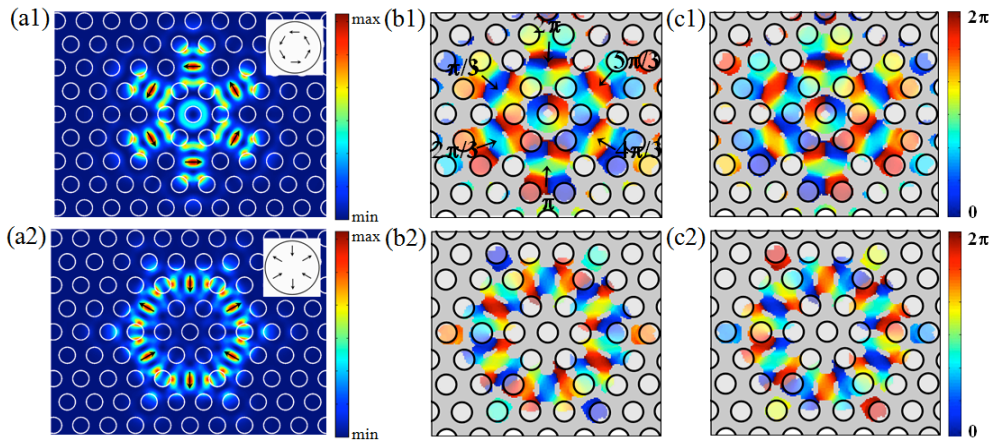

Fig.2. Near-field distributions of resonant Mode1 (top) and Mode2 (bottom). (a) Intensity distributions of the cavity modes, where the superimposed arrows denote polarization directions and the insets depict the polarization directions of the fields confined in the defects; (b)-(c) phase diagrams of the LH and RH circular polarization components of the cavity modes.

To analyze the vector characters of the cavity modes, we decompose them into left-hand (LH) and right-hand (RH) circular polarization components with phase distributions of $\delta_1$ and $\delta_2$, respectively. The optical fields of the two components can be described as $E_L = A_0\exp(i\delta_1)(\hat{e}_x - i\hat{e}_y)/2$ and $E_R = A_0\exp(i\delta_2)(\hat{e}_x + i\hat{e}_y)/2$, where $A_0$ is a constant factor, $\hat{e}_x$ and $\hat{e}_y$ are a pair of orthogonal base vectors in the Cartesian coordinate system [33]. The phase distributions of the LH and RH components are calculated, as shown in Figs. 2(b) and 2(c), where the top and bottom images correspond to Mode1 and Mode2, respectively. In the LH component of Mode1, the fields confined in the defects have different phase values [see Fig. 2(b1)]. Moreover, the phases increase anticlockwise in the step of $\pi/3$ among the defects, resulting in a $2\pi$-winding phase difference around the central air-hole. This phase distribution allows the formation of an optical vortex with topological charge of 1 via the far-field radiation [34], although the light distribution in near-field is discrete. The RH component of Mode1 possesses a clockwise increased $\pi/3$-step phase distribution, indicating the formation of a vortex beam with topological charge of -1 in the far-field. Similar step-phase distributions are also observed on the LH and RH components of Mode2 with opposite increased directions. However, different from the case of Mode1, the variation step among the defects is $2\pi/3$ with a $4\pi$-winding phase difference. Hence, the far-field radiation of the two components yields optical vortices with topological charges of 2 or -2, respectively. In the two modes, the phase distributions of LH and RH components satisfy $\delta_2 = -\delta_1$, and $\delta_1 = m\varphi$, where $m$ and $\varphi$ are the topological charge and the azimuthal angle to describe a helical phase structure, respectively. Therefore, the superposed field of the two components is derived as $E = A_0(\cos\delta\,\hat{e}_x + \sin\delta\,\hat{e}_y)$, which implies their superposition has vector polarization states [35].

*3.2 Generation of vector beams in a PPC cavity with N=6*

With above analysis and the results shown in Figs. 1(c) and 1(d), the polarization states and phase distributions of the resonant modes enable the combinations of vector beams in the scattered far-fields. To examine that, we numerically calculate the far-field radiation patterns of the cavity modes using the Sommerfeld diffraction integral

$$\mathbf{E}(x,y,z) = \frac{1}{2\pi} \nabla \times \iint \mathbf{n} \times \mathbf{E}_0(x_0, y_0, 0) \frac{\exp(ikR)}{R} dS \qquad (2)$$

where $E_0(x_0, y_0, 0)$ is the near-field in the PPC plane, $\nabla = \partial/\partial x \hat{e}_x + \partial/\partial y \hat{e}_y + \partial/\partial z \hat{e}_z$, and n is the unit vector in $z$ direction. Figure 3 shows the calculated far-fields in a transverse plane above the PPC slab, where the top and bottom images correspond to those of Mode1 and Mode2, respectively. Modulated by the symmetric polarization states and step-phase structures of the two resonant modes, the far-field emission patterns present ring- and petal-like intensity distributions, as shown in Figs. 3(a1) and 3(a2). The hollow-core intensity indicates the existence of a polarization singularity at the center [19].

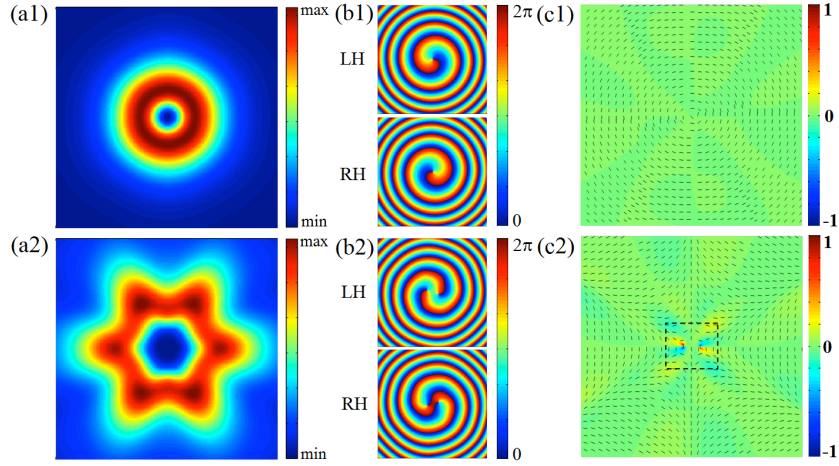

Fig.3. Numerical simulations of vector beams generated in far-field radiation of resonant Mode1 (top) and Mode2 (bottom). (a) Intensity distributions of the vector beams; (b) phase diagrams of LH and RH circular polarization components of the vector beams; (c) polarization states of the vector beams, where the background and short lines denote the ellipticity and orientation of major axis of polarization ellipse, respectively.

The phase distributions of the far-fields are then analyzed with their decomposed LH and RH circular polarization components, as shown in Figs. 3(b1) and 3(b2). The step-phase distributions of the resonant modes evolve into continuously helical structures and maintain the corresponding topological charges. However, unlike the standard vortex phase, the dislocation lines of the far-field phases are along helical tracks other than straight lines, which can be attributed to the appendix of a spherical wave during the diffraction of the scattered light. Similar to the analysis of vector characters of the near-field resonant modes, the far-fields are vector beams because of the opposite helical phase structures for the LH and RH components. Relying on the topological charges, we define the far-fields of Mode1 and Mode2 as the first- and second-order vector beams, respectively. The topological charge of 2 for the second-order vector beam results in a larger hollow-core in the intensity profile than that of the first-order vector beam.

We then study the polarization states of the generated vector beams by calculating their Stokes parameters [35]. The four Stokes parameters are defined as

$$\begin{aligned} S_0 &= E_x E_x^* + E_y E_y^* \\ S_1 &= E_x E_x^* - E_y E_y^* \\ S_2 &= E_x E_y^* + E_y E_x^* \\ S_3 &= i\left(E_x E_y^* - E_y E_x^*\right) \end{aligned} \quad (3)$$

where, $E_x$ and $E_y$ are the two polarization components along $x$- and $y$-axis, and the symbol "*" denotes the complex conjugation operation. With the Stokes parameters, an arbitrary polarization state can be described using the polarization ellipse. The orientation of the major axis $\psi$ and ellipticity tan$\chi$ of the ellipse are governed by

$$\tan 2\psi = S_2/S_1, \quad \sin\chi = S_3/S_0 \quad (4)$$

The ellipticity tan$\chi$=0, +1 and -1 correspond to linear polarization, RH and LH circular polarization components, respectively.

Figures 3(c1) and 3(c2) describe the polarization states of the first- and second-order vector beams, where the background and the short lines denote the ellipticity and orientation of major axis of polarization ellipse, respectively. As indicated in Figs. 3(c1)-3(c2), the ellipticity is calculated as zero, i.e., the both vector fields are locally linearly polarized. Note that, in the dashed box of Fig. 3(c2), the ellipticity distributions are discontinuous in the center region. The values of these discontinuous points have no physical meaning because the ellipticity of the polarization ellipse for the singularity is undefined. The polarization direction in Fig. 3(c1) is azimuthally varying, and that in Fig. 3(c2) has azimuthal and radial dependences, which shows first- and second-order vector character [3]. These results are consistent with results shown in Figs. 3(b1) and 3(b2).

*3.3 Improvement of extraction efficiency in a PPC cavity with N=6*

To implement the applications of the generated vector beams, such as in nano-lasers, single-photon sources, and cavity QED, it is desired to design the PPC cavity to have a moderately high extraction efficiency of the vector beams. Inspired by the improvement of off-chip coupling efficiency of cavity modes with perturbed air-holes [36,37], we enlarge the nearest neighbor air-holes around the defects. Here, the extraction efficiency $\eta$ is defined as the ratio of the collected optical power in the far-field to the total power of the cavity near-field. The perturbed PPC cavity is schematically shown in Fig. 4(a) with the radius ($r'$) of the perturbed air-holes increased from $0.3a$ to $0.32a$, which are denoted in red color. Figure 4(b) shows the near-field distribution of the resonant Mode1 for the perturbed cavity. Comparing with the result shown in Fig. 2(a1), we can confirm the perturbations do not break the symmetry of the cavity modes. In addition, the calculated far-field distribution of the perturbed cavity retains the intensity and polarization distributions demonstrated in Fig. 3. However, the extraction efficiency of Mode1 is improved from 1.83% for an unperturbed cavity to 8.82% for the perturbed one, while the $Q$ factor shows no observable degradation. The improvement of $\eta$ can be understood from the Fourier transform (FT) spectra of the resonant modes, as shown in Figs. 4(c) and 4(d) for Mode1 in the unperturbed and perturbed cavities, respectively. The red circles depict the light cone of the PPC slab, indicating the leaky regions of the resonant modes. Comparing with the result shown in Fig. 4(c), the FT spectrum in Fig. 4(d) contains large components inside the leaky region. Note that, by tuning positions and sizes of the perturbations, the coupling efficiencies and the $Q$ factors of the vector beams can be improved further [38, 39].

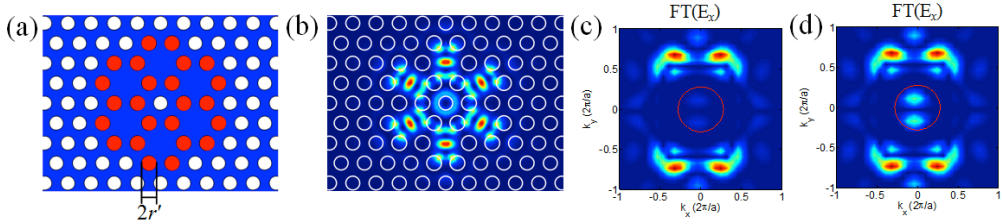

Fig. 4. (a) Schematic of the perturbed PPC cavity, where red air-holes are enlarged to have a radius of $r'=0.32a$; (b) near-field intensity distributions of the resonant Mode1 in the perturbed cavity; (c)-(d) FT spectra of Mode1 in the unperturbed and perturbed cavities.

*3.4 Vector beams generated in PPC cavities with N=3 and 4*

The generation of vector beams are also studied in multi-defect PPC cavities with $N=3$ and 4, as shown in Figs. 5(a) and 5(b), respectively. Where, the top and bottom images depict the near-field resonant modes and the corresponding far-field radiation patterns. Similar to the case of $N=6$, each PPC cavity presents two resonant modes with spatially dependent polarization distributions. In the defects, the optical fields have azimuthal or radial polarizations for the two resonant modes. Modulated by the polarizations of the near-field modes, hollow-core intensity distributions are observed in the far-fields. However, because of the lower symmetry for the cases of *N*=3 and 4, the far-field radiation have non-circular mode patterns. The vector characters of the far-fields are also confirmed by analyzing the phase distributions of LH and RH circular polarization components and the Stokes polarization parameters.

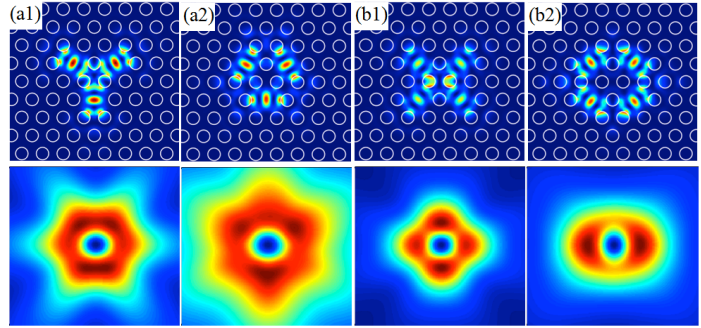

Fig.5. Near-field (top) and far-field (bottom) intensity distributions of resonant modes for PPC cavities with (a) *N*=3 and (b) *N*=4, where 1, 2 corresponding to Mode1 and Mode2, respectively.

## 4. Conclusions

In summary, we have demonstrated the generation of vector beams in PPCs by symmetrically missing multiple air-holes. With geometric symmetry of the defects relative to the cavity center, the resonant modes of PPC cavities show symmetric polarization states with respect to the central air-hole, and the vector characters are confirmed through the analysis of phase distributions of the decomposed LH and RH circular polarization components. Modulated with the polarization states and phase structures of the cavity modes, the far-field radiation patterns of the scattered light present vector beams with hollow-core intensity profiles. The vector characters of the far-field radiation patterns are analyzed carefully using the Stokes polarization parameters. In addition, the extraction efficiency of the vector beams from the PPC cavities could be improved by perturbing the air-holes around the defects. Comparing with other chip-integrated vector beam generators [8,9,20], the higher $Q/V_{mode}$ ratios of the PPC cavities promise the applications in on-chip micromanipulations, enhanced light-matter interactions and nonlinear processes involving the vector beams. The method demonstrated here may open up prospects for generating various structured lights in PPC cavities. The compact footprint of the PPC cavities promises the large-scale integration of such devices in a photonic integrated circuit.